\documentclass[a4paper, twocolumn]{article}
\usepackage{graphicx}
\usepackage{amsmath}

\newtheorem{definition}{Definition}
\newtheorem{theorem}{Theorem}
\begin{document}

\title{Entanglement bound for multipartite pure states based on local measurements }
\author{Li-zhen Jiang, Xiao-yu Chen ,Tian-yu Ye \\
{\small {College of Information and Electronic Engineering, Zhejiang
Gongshang University, Hangzhou, 310018, China}}}
\date{}
\maketitle

\begin{abstract}
An entanglement bound based on local measurements is introduced for
multipartite pure states. It is the upper bound of the geometric measure and
the relative entropy of entanglement. It is the lower bound of minimal
measurement entropy. For pure bipartite states, the bound is equal to the
entanglement entropy. The bound is applied to pure tripartite qubit states
and the exact tripartite relative entropy of entanglement is obtained for a
wide class of states.
\end{abstract}

\section{Introduction}

One of the open problem in quantum information theory is to quantify the
entanglement of a multipartite quantum state. Many entanglement measures for
pure or mixed multipartite states have been proposed \cite{Plenio} \cite
{Horodecki}, among them are the tangle \cite{Coffman} \cite{Miyake}, the
Schmidt measure \cite{Eisert} \cite{Hein} which is the logarithmic of the
minimal number of product terms that comprise the state vector, the
geometric measure \cite{Shimony} \cite{Wei1} \cite{Brody} which is defined
in terms of the maximal fidelity of the state vector and the set of pure
product states, the relative entropy of entanglement \cite{Vedral1} \cite
{Vedral2}, and the robustness \cite{Vidal} \cite{Harrow}. The last three are
related with each other \cite{Hayashi1} \cite{Hayashi2} \cite{Wei2} \cite
{Cavalcanti} \cite{Wei3} \cite{Zhu} and they are equal for some of the
states such as stabilizer states, symmetric basis and anti-symmetric basis
states. All these entanglement measures are not operationally defined. In
bipartite system, however, the entanglement cost and the distillable
entanglement are operational entanglement measures. If bipartite
entanglement measures satisfy some properties, it turns out that their
regularizations are bounded by distillable entanglement from one side and by
entanglement cost from the other side\cite{Horodecki}. For a pure bipartite
state, the two bounds are equal and the entanglement is simply the entropy
of the reduced density matrix thus has a clear information theoretical
meaning. We will investigate the possibility of extending these entropic and
operational definitions of entanglement to multipartite pure states in this
paper, we will propose an entanglement measurement bound (EMB) which is an
entanglement measure for pure tripartite qubit states. The bound is based on
the results of local measurements. Local measurement or local discrimination
had been used as upper bound of certain entanglement measures. For a graph
state, ''Pauli persistency'' has been used as an upper bound of Schmidt
measure \cite{Hein}, a quantity based on LOCC measurements has been used as
an upper bound \cite{Markham} of geometric measure.

\section{The justification of the definition of entanglement measurement
bound}

What is the usefulness of a bipartite entangled state? One answer should be
that we can use it for cryptography. If Alice's part is measured in spin up,
Bob's part should definitely in spin up for a bipartite spin entangled Bell
state $\Phi =\frac 1{\sqrt{2}}(\left| \uparrow \right\rangle _A\left|
\uparrow \right\rangle _B+\left| \downarrow \right\rangle _A\left|
\downarrow \right\rangle _B).$ Thus using $n$ pairs of Bell state, we can
get a shared string of $n$ bits for Alice and Bob through measurements. If
Alice measures her part in the basis of $\left| \phi \right\rangle =\cos
\phi \left| \uparrow \right\rangle +\sin \phi \left| \downarrow
\right\rangle ,$ $\left| \phi ^{\perp }\right\rangle =-\sin \phi \left|
\uparrow \right\rangle +\cos \phi \left| \downarrow \right\rangle ,$ then
the measurement results should be that Alice and Bob are simultaneously in
state $\left| \phi \right\rangle $ or they are simultaneously in state $%
\left| \phi ^{\perp }\right\rangle $ , each with probability $\frac 12.$
Thus if we have $n$ pairs of Bell state, we can get a shared string of $n$
bits regardless of the measurement basis Alice chosen. When we have a less
entangled state $\cos \theta \left| \uparrow \right\rangle _A\left| \uparrow
\right\rangle _B+\sin \theta \left| \downarrow \right\rangle _A\left|
\downarrow \right\rangle _B,$ what can we do for the purpose of
cryptography? Certainly, we can measure one of the part, say Alice in the
spin up and down basis. The result turns out to be a shared string of length
$n,$ with the probability $\cos ^2\theta $ for spin up, and the probability $%
\sin ^2\theta $ for spin down. The information contained in such a string
can be calculated to be $nH_2(\cos ^2\theta )$ , where $H_2(x)=-x\log
_2x-(1-x)\log _2(1-x)$ is the binary entropy function. However, if Alice
measures her part in a rather arbitrary basis $\left| \phi \right\rangle $
and $\left| \phi ^{\perp }\right\rangle ,$ Bob will get his state in $\left|
\phi _B\right\rangle \sim (\cos \phi \cos \theta \left| \uparrow
\right\rangle _B+\sin \phi \sin \theta \left| \downarrow \right\rangle _B)$
and $\left| \phi _B^{\perp }\right\rangle \sim (-\sin \phi \cos \theta
\left| \uparrow \right\rangle _B+\cos \phi \sin \theta \left| \downarrow
\right\rangle _B)$ states with probabilities $p(\theta ,\phi )=\cos ^2\theta
\cos ^2\phi +\sin ^2\theta \sin ^2\phi $ and $1-p(\theta ,\phi )$,
respectively. We can get a shared string of length $n$ with total
information content $nH_2(p(\theta ,\phi )).$ Notice that $p(\theta ,\phi
)=\cos 2\theta \cos ^2\phi +\sin ^2\theta ,$ the minimal information content
occurs at the case of $\cos ^2\phi =1$ or $0.$ So at least we can get a
shared string with information content $nH_2(\cos ^2\theta ),$ which is the
entanglement of the $n$ pairs of the state $\cos \theta \left| \uparrow
\right\rangle _A\left| \uparrow \right\rangle _B+\sin \theta \left|
\downarrow \right\rangle _A\left| \downarrow \right\rangle _B.$ Hence, the
entanglement of a pure bipartite entangled state is the minimal shared
information content obtained by measurement. This point had been proved in
Ref. \cite{Chernyavskiy} in the context of measurement entropy. For
completeness, we will give an alternative proof in the following.

For a bipartite state $\left| \psi \right\rangle =\sum_{i,j=1}^dA_{ij}\left|
i\right\rangle \left| j\right\rangle $ , where $\left| i\right\rangle $ and $%
\left| j\right\rangle $ are the orthogonal basis, it is well known that the
entanglement entropy is the entropy of the reduced density matrix when one
of the partite is traced out. We have $\rho _A=Tr_B(\left| \psi
\right\rangle \left\langle \psi \right| )=\sum_{i,j,k}A_{ij}A_{kj}^{*}\left|
i\right\rangle \left\langle k\right| .$ Thus in the basis $\left|
i\right\rangle $, the reduced state is $\rho _A=AA^{\dagger },$where $A$ is
the matrix with entries $A_{ij}.$ A unitary transformation $U$ diagonalizes $%
\rho _A$ to $\Lambda =diag\{\lambda _1,\ldots ,\lambda _d\},$ thus $%
AA^{\dagger }=U\Lambda U^{\dagger }.$ So that the singular value
decomposition of matrix $A$ is $A=U\sqrt{\Lambda }V^T,$ where $V$ is some
other unitary transformation. We have the Schmidt decomposition
\begin{equation}
\left| \psi \right\rangle =\sum_k\sqrt{\lambda _k}\left| \varphi
_k^{(1)}\right\rangle \left| \varphi _k^{(2)}\right\rangle ,  \label{wav2}
\end{equation}
with orthogonal basis$\left| \varphi _k^{(1)}\right\rangle
=\sum_iU_{ik}\left| i\right\rangle ,$ $\left| \varphi _k^{(2)}\right\rangle
=\sum_jV_{jk}\left| j\right\rangle .$ The entanglement of the state is
\begin{equation}
E(\left| \psi \right\rangle )=-\sum_{i=1}^d\lambda _k\log _2\lambda _k.
\label{wav3}
\end{equation}
Now, we use the measurement to obtain shared digital information from the
state vector $\left| \psi \right\rangle =\sum_{i,j=1}^dA_{ij}\left|
i\right\rangle \left| j\right\rangle .$ Suppose the state vector is
projected to Alice's measurement base vector $\left| c\right\rangle
=\sum_i^dc_i\left| i\right\rangle ,$ then Bob's state will be proportional
to $\left\langle c\right. \left| \psi \right\rangle
=\sum_{i,j}c_i^{*}A_{ij}\left| j\right\rangle ,$ the probability of which is
\begin{equation}
p=\sum_j\left| \sum_ic_i^{*}A_{ij}\right| ^2.
\end{equation}
Let's consider which measurement basis yields the optimal probability $p.$
This is an optimal of $p$ with respect to $\{c_i\}$ subjected to $%
\sum_i\left| c_i\right| ^2=1.$ With the Lagrange multiplier $\lambda $, we
can write the optimal equation as $\frac{\partial L}{\partial c_i^{*}}=0$,
where $L=p-\lambda (\sum_i\left| c_i\right| ^2-1).$ The optimal equation
then reads
\begin{equation}
\sum_{i,j}A_{ij}A_{kj}^{*}c_k-\lambda c_k=0,
\end{equation}
or
\[
(AA^{\dagger }-\lambda )\mathbf{c=}0,
\]
where $\mathbf{c=}(c_1,\ldots ,c_d)^T.$ The optimal probability should be $%
p=\sum_{i,j}c_i^{*}A_{ij}A_{kj}^{*}c_k$ $=\mathbf{c}^{\dagger }AA^{\dagger }%
\mathbf{c}$ $\mathbf{=c}^{\dagger }\lambda \mathbf{c}=\lambda .$ So the
optimal probability is the eigenvalue of the reduced density matrix $\rho
_A=AA^{\dagger }.$ Hence if we use eigenvectors of $\rho _A$ as the
measurement basis, the average information of each shared digit is the
entanglement of the state. Denote the eigensystem of $\rho _A$ as $\{\lambda
_k,\mathbf{c}^k\},$ let's see if the unitary transformed basis $\{U\mathbf{c}%
^k\}$ decrease the entropy $H(\mathbf{p)=-}\sum_{i=1}^dp_k\log _2p_k$ or
not, where $p_k=\mathbf{c}^{k\dagger }U^{\dagger }AA^{\dagger }U\mathbf{c}%
^k. $ Denote the elements of $U$ in the basis of $\mathbf{c}^k$, we have $%
U_{ij}=\mathbf{c}^{i\dagger }U\mathbf{c}^j$. Using the spectrum
decomposition of $AA^{\dagger },$ then
\begin{eqnarray}
p_k &=&\sum_i\lambda _i\mathbf{c}^{k\dagger }U^{\dagger }\mathbf{c}^i\mathbf{%
c}^{i\dagger }U\mathbf{c}^k  \nonumber \\
&=&\sum_i\lambda _i\left| U_{ik}\right| ^2.
\end{eqnarray}
Notice that function $f(x)=-x\log _2x$ is concave, that is, for $\alpha \in
[0,1],$ one has $f(\alpha x_1+(1-\alpha )x_2)\geq \alpha f(x_1)+(1-\alpha
)f(x_2).$ Then $f(p_k)\geq \sum_i\left| U_{ik}\right| ^2f(\lambda _i),$
where the unitarity of $U$ is used. Hence
\begin{eqnarray}
H(\mathbf{p)} &=&\sum_kf(p_k)\geq \sum_{i,k}\left| U_{ik}\right| ^2f(\lambda
_i)  \nonumber \\
&=&\sum_if(\lambda _i)=E(\left| \psi \right\rangle ).
\end{eqnarray}
We get the desired result.

\subsection{Definition}

\begin{definition}
For an $n-$partite state $\left| \psi \right\rangle ,$ let $\mathbf{p}$ be
the probability vector with multiple subscripts, the components of $\mathbf{p%
}$ are $p_{i_1,i_2,\ldots ,i_N}=$ $\left| \left\langle \phi
_{i_1}^{(1)}\right| \otimes \left\langle \phi _{i_2|i_1}^{(2)}\right|
\otimes \cdots \left\langle \phi _{i_N|i_1i_2\ldots i_N}^{(N)}\right| \text{
}\cdot \left| \psi \right\rangle \right| ^2.$ Here $\left| \phi
_{i_j|i_1i_2\ldots i_{j-1}}^{(j)}\right\rangle $ ( denoted simply as $\left|
\phi _{i_j}^{(j)}\right\rangle $ hereafter) $(i_j=0,1,\ldots ,d_j-1)$ are
the orthonormal basis of $j-th$ partite when the measurement results for the
former parties are $i_1,i_2,\ldots ,i_{j-1},$ respectively. The EMB of $%
\left| \psi \right\rangle $ is defined as the minimal entropy of the
measurement probability vector, that is,
\begin{equation}
E_{MB}(\left| \psi \right\rangle )=\min_{\mathbf{p}}H(\mathbf{p).}
\label{wav4}
\end{equation}
The minimization is over all possible local orthogonal measurements.
\end{definition}

Notice that the local measurements can be carried out step by step, for each
result of the first partite measurement, one can choose an orthogonal basis
to measure the second partite residue state. So one may have $d_1$ different
projection measurements for the second partite. When measuring the third
partite, one can have $d_1d_2$ projection measurements, and so on. The
choice of $j-th$ partite basis can rely on all the former measurement
results. The total number of measurements is $1+d_1+d_1d_2+\ldots
+d_1d_2\cdots d_{N-1}.$ Meanwhile, the minimization in (\ref{wav4}) is also
with respect to all permutation of the parties.

There is an definition of entanglement measure based on measurement\cite
{Chernyavskiy}. In the definition of \cite{Chernyavskiy}, each partite has
only one kind of (complete) measurement, the total number of the
measurements is $N-1.$ The $E_{Hmin}$ in \cite{Chernyavskiy} is no less than
our entanglement measurement bound $E_{MB}(\left| \psi \right\rangle )$ by
definition.

\section{As upper bounds of entanglement measures}

\subsection{Coarse grain}

In the multipartite case it is useful to compare EMB according to different
partitions, where the components $1,...,N$ are grouped into disjoint sets.
Any sequence $(A_1,...,A_N)$ of disjoint subsets $A\in V$ with $%
\bigcup_{i=1}^NA_i=\{1,...,N\}$ will be called a partition of $V$ . We will
write
\begin{equation}
(A_1,...A_N)\leq (B_1,...,B_M),
\end{equation}
if $(A_1,...A_N)$ is a finer partition than $(B_1,...,B_M)$. EMB is
non-increasing under a coarser grain of the partition. If two components are
merged to form a new component, then EMB can only decrease. This is because
that the minimization in the definition of EMB Eq. (\ref{wav4}) can also be
seen as with respect to all possible local measurement hierarchies. A local
measurement hierarchy of a finer partition $(A_1,...A_N)$ is definitely a
local measurement hierarchy of the coarser grain partition $(B_1,...,B_M),$
while the inverse may not be true. So from $(A_1,...A_N)$ to $(B_1,...,B_M),$
the set of local measurement hierarchies is enlarged, the minimization may
reach further lower value. We have
\begin{equation}
E_{MB}^{(A_1,...A_N)}(\left| \psi \right\rangle )\geq
E_{MB}^{(B_1,...B_M)}(\left| \psi \right\rangle ),  \label{wav5}
\end{equation}
where we specify the partition as the superscript of EMB. So any coarser
partition is a lower bound of the finer partition for EMB. Especially, lower
bound of EMB for a tripartite pure state is the bipartite pure state
entanglement, which is easily obtained. There are three bipartitions of a
tripartite state, the tighter lower bound is the partition with largest
entanglement.

\subsection{Upper bound of geometric measure}

Suppose $E_{MB}(\left| \psi \right\rangle )$ is achieved by the probability
vector $\mathbf{p}$ with components $p_{i_1,i_2,\ldots ,i_n}.$ For all $%
p_{i_1,i_2,\ldots ,i_n}=\left| \left\langle \phi _{i_1}^{(1)}\right| \otimes
\left\langle \phi _{i_2}^{(2)}\right| \otimes \cdots \left\langle \phi
_{i_N}^{(N)}\right| \text{ }\cdot \left| \psi \right\rangle \right| ^2,$ we
may denote the largest one as $p_{0,0,\ldots ,0}=\left| \left\langle \phi
_0^{(1)}\right| \otimes \left\langle \phi _0^{(2)}\right| \otimes \cdots
\left\langle \phi _0^{(N)}\right| \text{ }\cdot \left| \psi \right\rangle
\right| ^2.$ So $p_{0,0,\ldots ,0}\geq p_{i_1,i_2,\ldots ,i_N}$. Then $%
p_{i_1,i_2,\ldots ,i_n}\log _2p_{i_1,i_2,\ldots ,i_N}\leq p_{i_1,i_2,\ldots
,i_N}\log _2p_{0,0,\ldots ,0}$
\begin{eqnarray}
E_{MB}(\left| \psi \right\rangle ) &=&-\sum_{i_1,i_2,\ldots
,i_N}p_{i_1,i_2,\ldots ,i_N}\log _2p_{i_1,i_2,\ldots ,i_N}  \nonumber \\
&\geq &-\sum_{i_1,i_2,\ldots ,i_N}p_{i_1,i_2,\ldots ,i_N}\log
_2p_{0,0,\ldots ,0}  \nonumber \\
&\geq &-\log _2p_{0,0,\ldots ,0}  \nonumber \\
&\geq &E_G(\left| \psi \right\rangle ).
\end{eqnarray}
The last inequality comes from the fact that the geometric measure $%
E_G(\left| \psi \right\rangle )=\min -\log _2F,$ where $F=\left|
\left\langle \varphi ^{(1)}\right| \otimes \left\langle \varphi
^{(2)}\right| \otimes \cdots \left\langle \varphi ^{(N)}\right| \text{ }%
\cdot \left| \psi \right\rangle \right| ^2,$ the minimization is over all
possible product state $\left| \varphi ^{(1)}\right\rangle \otimes \left|
\varphi ^{(2)}\right\rangle \otimes \cdots \left| \varphi
^{(N)}\right\rangle .$ The largest fidelity $F$ should be no less than some
special fidelity $p_{0,0,\ldots ,0}.$

\subsection{Upper bound of the relative entropy of entanglement}

Suppose the orthogonal expansion of $\left| \psi \right\rangle
=\sum_{i_1,i_2,\ldots ,i_N}\xi _{i_1,i_2,\ldots ,i_N}\left| \phi
_{i_1}^{(1)}\right\rangle \otimes \left| \phi _{i_2}^{(2)}\right\rangle
\otimes \cdots \left| \phi _{i_N}^{(N)}\right\rangle $ with $\xi
_{i_1,i_2,\ldots ,i_N}=\left\langle \phi _{i_1}^{(1)}\right| \otimes
\left\langle \phi _{i_2}^{(2)}\right| \otimes \cdots \left\langle \phi
_{i_N}^{(N)}\right| $ $\cdot \left| \psi \right\rangle $ be the optimal
expansion that achieves the measure entanglement, that is $E_{MB}(\left|
\psi \right\rangle )=-\sum_{i_1,i_2,\ldots ,i_N}p_{i_1,i_2,\ldots ,i_N}\log
_2p_{i_1,i_2,\ldots ,i_N}$ with $p_{i_1,i_2,\ldots ,i_N}=\left| \xi
_{i_1,i_2,\ldots ,i_N}\right| ^2.$ Let's construct the separable state
\begin{eqnarray}
\omega &=&\sum_{i_1,i_2,\ldots ,i_N}p_{i_1,i_2,\ldots ,i_N}\left| \phi
_{i_1}^{(1)}\right\rangle \left| \phi _{i_2}^{(2)}\right\rangle \cdots
\left| \phi _{i_N}^{(N)}\right\rangle  \nonumber \\
&&\times \left\langle \phi _{i_1}^{(1)}\right| \left\langle \phi
_{i_2}^{(2)}\right| \cdots \left\langle \phi _{i_N}^{(N)}\right| .
\end{eqnarray}
The relative entropy of $\left| \psi \right\rangle $ with respect to $\omega
$ is $-Tr\left| \psi \right\rangle \left\langle \psi \right| \log _2\omega =$
$-\left\langle \psi \right| \log _2\omega \left| \psi \right\rangle
=-\left\langle \psi \right| \sum_{i_1,i_2,\ldots ,i_N}\left| \phi
_{i_1}^{(1)}\right\rangle \left| \phi _{i_2}^{(2)}\right\rangle \cdots
\left| \phi _{i_N}^{(N)}\right\rangle $ $\log _2p_{i_1,i_2,\ldots ,i_N}$ $%
\left\langle \phi _{i_1}^{(1)}\right| \left\langle \phi _{i_2}^{(2)}\right|
\cdots \left\langle \phi _{i_N}^{(N)}\right| \left. \psi \right\rangle $ $%
=E_{MB}(\left| \psi \right\rangle ).$ It is larger than or equal to the
relative entropy of entanglement $E_R(\left| \psi \right\rangle ).$ Since
the separable state $\omega $ is just one of the full separable states, it
may not be the full separable state that achieves the minimal relative
entropy for state $\left| \psi \right\rangle .$ So we have
\begin{equation}
E_{MB}(\left| \psi \right\rangle )\geq E_R(\left| \psi \right\rangle ).
\end{equation}

More concretely, we will consider the pure tripartite qubit state in the
next section.

\section{Pure tripartite qubit state}

It is well known that GHZ state and W state are two different kinds of pure
tripartite states that are not convertible with each other under stochastic
local operation and classical communication (SLOCC). We may write the states
in computational basis as $\left| GHZ\right\rangle =\frac 1{\sqrt{2}}(\left|
000\right\rangle +\left| 111\right\rangle )$ and $\left| W\right\rangle
=\frac 1{\sqrt{3}}(\left| 001\right\rangle +\left| 010\right\rangle +\left|
100\right\rangle ).$ The three parties are called Alice, Bob and Charlie.
When they share GHZ state, if Alice measures her part with result $0,$ then
the states of Bob and Charlie are in $0$ without further measurement. When
Alice measures $1,$ the other two parts are also in $1.$ A common string of
bits among the three parts can be established when one of them measures in
computational basis. However, when Alice measures in $\left| \phi
\right\rangle =\cos \phi \left| 0\right\rangle +\sin \phi \left|
1\right\rangle ,$ $\left| \phi ^{\perp }\right\rangle =-\sin \phi \left|
0\right\rangle +\cos \phi \left| 1\right\rangle $ basis, the joint state of
Bob and Charlie will be left to the entangled state $\cos \phi \left|
00\right\rangle +\sin \phi \left| 11\right\rangle $ or $-\sin \phi \left|
00\right\rangle +\cos \phi \left| 11\right\rangle $. Further measurement
should be performed to determine the state of Bob as well as Charlie. So GHZ
state measured in computational basis is a rather special case when only one
step of measurement is required to transform the tripartite entanglement to
shared bits. In general, we need two steps of measurements to convert the
tripartite quantum correlation to classical correlation.

In computational basis, a pure tripartite qubit state can be written as
\begin{equation}
\left| \psi \right\rangle =\sum_{i,j,k=0}^1A_{ijk}\left| i\right\rangle
\left| j\right\rangle \left| k\right\rangle ,  \label{wav6}
\end{equation}
the normalization takes $\sum_{i,j,k=0}^1\left| A_{ijk}\right| ^2=1.$ Let
the measurement basis of Alice are $\left| \phi _a\right\rangle
=a_0^{*}\left| 0\right\rangle +a_1^{*}\left| 1\right\rangle ,$ $\left| \phi
_a^{\perp }\right\rangle =-a_1\left| 0\right\rangle +a_0\left|
1\right\rangle ,$ the basis of Bob are $\left| \phi _b\right\rangle
=b_0^{*}\left| 0\right\rangle +b_1^{*}\left| 1\right\rangle ,$ $\left| \phi
_b^{\perp }\right\rangle =-b_1\left| 0\right\rangle +b_0\left|
1\right\rangle $ when Alice is projected to $\left| \phi _a\right\rangle $,
the basis of Bob are $\left| \phi _b^{\prime }\right\rangle =b_0^{\prime
*}\left| 0\right\rangle +b_1^{\prime *}\left| 1\right\rangle ,$ $\left| \phi
_b^{\prime \perp }\right\rangle =-b_1^{\prime }\left| 0\right\rangle
+b_0^{\prime }\left| 1\right\rangle $ when Alice is projected to $\left|
\phi _a^{\perp }\right\rangle $. Suppose the state $\left| \psi
\right\rangle $ be projected to $\left| \phi _a\right\rangle \left| \phi
_b\right\rangle $ for Alice and Bob's parts, then Charlie should be left in $%
\left\langle \phi _a\phi _b\right| \left. \psi \right\rangle $ $=$ $%
\sum_{k=0}^1(\sum_{i,j=0}^1A_{ijk}a_ib_j)\left| k\right\rangle $ , the
probability of measurement is $p_{ab}=\sum_{k=0}^1\left|
\sum_{i,j=0}^1A_{ijk}a_ib_j\right| ^2$. We may write $p_{ab}=p_ap_{b|a},$
where $p_a$ is the probability of projecting $\left| \psi \right\rangle $ to
$\left| \phi _a\right\rangle ,$ and $p_{b|a}$ is the probability of
projecting further to $\left| \phi _b\right\rangle .$ For all possible local
measurements, we consider the minimal entropy of the probability
distribution $\{p_{ab},p_{ab^{\perp }},p_{a^{\perp }b^{\prime }},p_{a^{\perp
}b^{\prime \perp }}\},$ alternatively, we may write it as $\{p_{00},$ $%
p_{01},p_{10},p_{11}\}.$ The entropy of the measurement should be
\begin{eqnarray*}
E_{MB}(\left| \psi \right\rangle ) &=&\min (-\sum_{i,j=0}^1p_{ij}\log p_{ij})
\\
&=&\min [-\sum_i^1(p_i\log p_i+p_i\sum_jp_{j|i}\log p_{j|i}].
\end{eqnarray*}
So we may solve the problem by minimizing the entropy of conditional
distribution $p_{j|i}$ by first fixing $p_i,$ that is, after Alice's part is
measured in some basis that is not known to Bob and Charlie, Bob choose some
basis to minimize the entropy of conditional distribution $p_{j|i}.$ Since
the joint state of Bob and Charlie is left to (unnormalized) $\left\langle
\phi _a\right. \left| \psi \right\rangle =\sum_{i,j,k=0}^1A_{ijk}a_i\left|
j\right\rangle \left| k\right\rangle $ for some quite general measurement
base $\left| \phi _a\right\rangle $ of Alice. From the result of bipartite
case, we have
\begin{equation}
\min -\sum_{j=0}^1p_{j|i}\log p_{j|i}=E(\left| \psi _i\right\rangle ).
\end{equation}
Thus the minimization problem turns out to be
\begin{equation}
E_{MB}(\left| \psi \right\rangle )=\min_{\{a_0,a_1\}}[-\sum_{i=0}^1(p_i\log
p_i+p_iE(\left| \psi _i\right\rangle )].  \label{wav7}
\end{equation}
Where
\begin{eqnarray}
\left| \psi _0\right\rangle &=&p_0^{-1/2}\sum_{i,j,k=0}^1A_{ijk}a_i\left|
j\right\rangle \left| k\right\rangle ,  \label{wav7a} \\
\left| \psi _1\right\rangle
&=&p_1^{-1/2}\sum_{j,k=0}^1(-A_{0jk}a_1^{*}+A_{1jk}a_0^{*})\left|
j\right\rangle \left| k\right\rangle ;  \label{wav7b}
\end{eqnarray}
with
\begin{eqnarray}
p_0 &=&\sum_{j,k=0}^1\left| A_{0jk}a_0+A_{1jk}a_1\right| ^2,  \label{wav7c}
\\
p_1 &=&\sum_{j,k=0}^1\left| -A_{0jk}a_1^{*}+A_{1jk}a_0^{*}\right| ^2.
\label{wav7d}
\end{eqnarray}
In practical calculation, we can choose $a_0=\cos \theta ,$ $a_1=\sin \theta
e^{i\varphi },$ thus $E_{MB}(\left| \psi \right\rangle )$ is given by the
minimization over $\left\{ \theta ,\varphi \right\} $. The bipartite
entanglement at RHS of (\ref{wav7}) can easily be evaluated with
concurrence. Alternatively, we may write EMB as
\begin{equation}
E_{MB}(\left| \psi \right\rangle
)=\min_{\{a_0,a_1\}}[\sum_{i=0}^1S(B_iB_i^{\dagger })],  \label{wav8}
\end{equation}
where $S$ is the von Neumann entropy of a matrix, $S(\varrho )=-Tr(\varrho
\log _2\varrho ),$
\begin{eqnarray}
B_0 &=&a_0A_0+a_1A_1, \\
B_1 &=&-a_1^{*}A_0+a_0^{*}A_1,
\end{eqnarray}
with $A_0$ and $A_1$ are the matrices of elements $\left( A_0\right)
_{jk}=A_{0jk},$ $\left( A_1\right) _{jk}=A_{1jk}.$

For $E_{Hmin}$ in \cite{Chernyavskiy}, Bob's measurement is independent of
Alice's measurements. Suppose the measurement basis of Alice be $\left| \phi
_a\right\rangle =a_0^{*}\left| 0\right\rangle +a_1^{*}\left| 1\right\rangle
, $ $\left| \phi _a^{\perp }\right\rangle =-a_1\left| 0\right\rangle
+a_0\left| 1\right\rangle ,$ the basis of Bob be $\left| \phi
_b\right\rangle =b_0^{*}\left| 0\right\rangle +b_1^{*}\left| 1\right\rangle
, $ $\left| \phi _b^{\perp }\right\rangle =-b_1\left| 0\right\rangle
+b_0\left| 1\right\rangle $, respectively. Then
\begin{equation}
E_{Hmin}(\left| \psi \right\rangle )=\min (-\sum_{i,j=0}^1p_{ij}^{\prime
}\log p_{ij}^{\prime }).  \label{wav8a}
\end{equation}
where $p_{lm}^{\prime }=\sum_{k=0}^1\left|
\sum_{ij}A_{ijk}a_{li}b_{mj}^{}\right| ^2,$ with $%
(a_{00},a_{01},a_{10},a_{11})=(a_0,a_1,-a_1^{*},a_0^{*})$ and $%
(b_{00},b_{01},b_{10},b_{11})=(b_0,b_1,-b_1^{*},b_0^{*}).$ The
minimization in (\ref{wav8a}) is more difficult than that of
(\ref{wav8}). The calculation of the geometric measure involves
minimization over the product state of three qubits and thus is
more difficult than the calculation of EMB in (\ref{wav8}). Only
for symmetric tripartite state, the calculation of the geometric
measure can be reduced as shown later.

\subsection{Superposition of GHZ and W' states}

It has been known \cite{Carteret} \cite{Tamaryan} that any pure tripartite
qubit state can be local unitarily transformed to the standard form
\begin{eqnarray}
\left| \psi \right\rangle &=&q_0\left| 000\right\rangle +q_1\left|
011\right\rangle +q_2\left| 101\right\rangle  \nonumber \\
&&+q_3\left| 110\right\rangle +q_4e^{i\gamma }\left| 111\right\rangle .
\label{wav9}
\end{eqnarray}
Where $q_i$ are positive, $\gamma \in [-\pi /2,\pi /2]$. The concurrences of
$\left| \psi _0\right\rangle $ and $\left| \psi _1\right\rangle $ are $%
C_0=2p_0^{-1}\left| q_0q_1\cos ^2\theta +\text{ }q_0q_4\sin \theta \cos
\theta e^{i(\gamma +\varphi )}\text{ }-q_2q_3\sin ^2\theta e^{2i\varphi
}\right| ,$ with probability $p_0=(q_0^2+q_1^2)\cos ^2\theta
+(q_2^2+q_3^2+q_4^2)\sin ^2\theta +2q_1q_4\sin \theta \cos \theta \cos
(\gamma +\varphi )$, and $C_1=$ $2p_1^{-1}\left| q_0q_1\sin ^2\theta -\text{
}q_0q_4\sin \theta \cos \theta e^{i(\gamma +\varphi )}\text{ }-q_2q_3\cos
^2\theta e^{2i\varphi }\right| ,$ with probability $p_1=(q_0^2+q_1^2)\sin
^2\theta +(q_2^2+q_3^2+q_4^2)\cos ^2\theta -2q_1q_4\sin \theta \cos \theta
\cos (\gamma +\varphi )$, respectively.

A special case is the superposition of GHZ and W' state, $\left|
GHZ-W^{\prime }\right\rangle =\cos \alpha \left| GHZ\right\rangle +\sin
\alpha \left| W^{\prime }\right\rangle $, which is a standard tripartite
state with $q_0=q_4=\frac 1{\sqrt{2}}\cos \alpha ,$ $q_1=q_2$ $=q_3=\frac 1{%
\sqrt{3}}\sin \alpha ,$ $\gamma =0.$ The state is widely used in evaluating
the tangle of symmetric tripartite mixed state. For this superposition
state, we have calculated EMB for parameter $x=\sin \alpha $, $\alpha \in
[-\pi /2,\pi /2].$ The results are shown in Fig.1. Also shown in Fig.1 are
the tangle, the geometric measure and the bipartition entanglement of the
state. The tangle of the state is \cite{Eltschka}
\[
\tau (\left| GHZ-W^{\prime }\right\rangle )=\left| \cos ^4\alpha +\frac 89%
\sqrt{6}\sin ^3\alpha \cos \alpha \right| .
\]
According to the permutation symmetry of the state, the geometric measure
for this state is \cite{Hubener} \cite{Hayashi3} \cite{Wei4}
\begin{equation}
E_G=\min_\phi -\log _2\left| \left\langle GHZ-W^{\prime }\right| \left(
\left| \phi \right\rangle \right) ^{\otimes 3}\right| ^2,  \label{wav8b}
\end{equation}
where $\left| \phi \right\rangle $ is a qubit state.

\begin{figure}[tbp]
\includegraphics[ trim=0.000000in 0.000000in -0.138042in 0.000000in,
height=2.0081in, width=2.5097in]{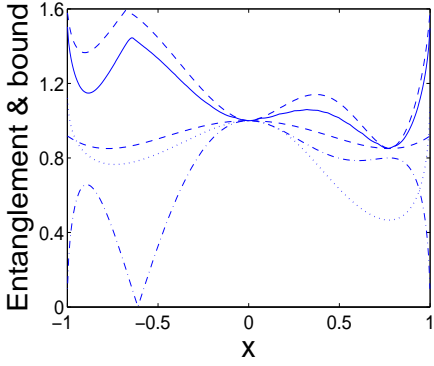}
\par
\vspace{5mm} \caption{The Entanglement with respect to x, the
portion of W' state in the superposition of W' and GHZ state. The
solid line is the entanglement measurement bound, the dotted line
is the geometric measure, the up dashed line is $E_{Hmin}$ of Ref.
\protect\cite{Chernyavskiy}, the down dashed line is the bipartite
entanglement, the dash-dot line is the tangle.}
\end{figure}

\subsection{Bipartite lower bound}

For a general state $\left| \psi \right\rangle
=\sum_{i,j,k=0}^1A_{ijk}\left| i\right\rangle \left| j\right\rangle \left|
k\right\rangle ,$ we may project it to state $\left| \phi _{ab}\right\rangle
=\sum_{i,j=0}^1c_{ij}^{*}\left| i\right\rangle \left| j\right\rangle $ with
joint measurement of Alice and Bob, then Charlie should be left in $%
\left\langle \phi _a{}_b\right| \left. \psi \right\rangle $ $=$ $%
\sum_{k=0}^1(\sum_{i,j=0}^1A_{ijk}c_i{}_j)\left| k\right\rangle $. The
bipartition is a coarser grain of a tripartition, so
\begin{equation}
E_{bi}(\left| \psi \right\rangle )\leq E_{MB}(\left| \psi \right\rangle ).
\end{equation}
The bipartite lower bound is $E_{bi}(\left| \psi \right\rangle )=\min
\{H_2(x_1),H_2(x_2),H_2(x_3)\}$, with $x_m=\frac 12(1+\sqrt{1-C_m^2}).$ The
concurrence $C_m=2\sqrt{\left|
d_{00}^{(m)}d_{11}^{(m)}-d_{01}^{(m)}d_{10}^{(m)}\right| },$ with $%
d_{ii^{\prime }}^{(1)}=\sum_{j,k=0}^1A_{ijk}A_{i^{\prime }jk}^{*},$ $%
d_{jj^{\prime }}^{(2)}=\sum_{i,k=0}^1A_{ijk}A_{ij^{\prime }k}^{*},$ $%
d_{kk^{\prime }}^{(3)}=\sum_{i,j=0}^1A_{ijk}A_{ijk^{\prime }}^{*}.$ For a $%
\left| GHZ-W^{\prime }\right\rangle $ state, the three concurrences are
equal to
\begin{equation}
C=\sqrt{\cos ^4\alpha +\frac 43\sin ^2\alpha \cos ^2\alpha +\frac 89\sin
^4\alpha }.
\end{equation}
So the lower bound of EMB of $\left| GHZ-W^{\prime }\right\rangle $ state is
$H_2(\frac 12(1+\sqrt{1-C^2})).$

\subsection{A special superposition state with equal tripartite EMB and
bipartite entanglement}

It can be seen from figure 1 that there is a superposition of GHZ and W'
state whose tripartite EMB and bipartite entanglement are equal. The state
is a $\left| GHZ-W^{\prime }\right\rangle $ state with $x=\sin \alpha =\sqrt{%
\frac 35},$ we will denote it as $\left| \Omega \right\rangle $ in the
following. Then
\begin{eqnarray}
\left| \Omega \right\rangle &=&\frac 1{\sqrt{5}}(\left| 000\right\rangle
+\left| 011\right\rangle +\left| 101\right\rangle  \nonumber \\
&&+\left| 110\right\rangle +\left| 111\right\rangle ).
\end{eqnarray}
The bipartite entanglement is $E_{bi}(\left| \Omega \right\rangle )=S(\rho
_C),$ where $\rho _C=Tr_{AB}(\left| \Omega \right\rangle \left\langle \Omega
\right| )=\frac 25\left| 0\right\rangle \left\langle 0\right| +\frac
15\left| 0\right\rangle \left\langle 1\right| +\frac 15\left| 1\right\rangle
\left\langle 0\right| +\frac 35\left| 1\right\rangle \left\langle 1\right| .$
Then
\begin{equation}
E_{bi}(\left| \Omega \right\rangle )=H_2[\frac 12(1+\frac 1{\sqrt{5}%
})]\approx 0.8505.
\end{equation}
For the tripartite EMB, the eigenvalues of $B_0B_0^{\dagger }$ and $%
B_1B_1^{\dagger }$ in Eq.(\ref{wav8}) are
\begin{eqnarray}
\lambda _{0\pm } &=&\frac 1{10}(2+K\pm \sqrt{5}K), \\
\lambda _{1\pm } &=&\frac 1{10}[3-K\pm \sqrt{5}(1-K)],
\end{eqnarray}
respectively, where $K=\left| a_1\right| ^2+a_0a_1^{*}+a_1a_0^{*}=\sin
^2\theta +\sin 2\theta \cos \varphi .$ Notice that $\lambda _{0+}+\lambda
_{1-}=\frac 1{10}(5+\sqrt{5}),\lambda _{0-}+\lambda _{1+}=\frac 1{10}(5-%
\sqrt{5}),$ the minimal entropy summation in Eq.(\ref{wav8}) should be
achieved by maximal $K$ or minimal $K.$ The maximal and minimal values of $K$
are $\frac 12(1\pm \sqrt{5}),$ respectively. Either of them leads to the
same eigenvalues $\{0,0,$ $\frac 12(1+\frac 1{\sqrt{5}}),\frac 12(1-\frac 1{%
\sqrt{5}})$ $\}.$ The tripartite EMB then is
\begin{equation}
E_{MB}(\left| \Omega \right\rangle )=H_2[\frac 12(1+\frac 1{\sqrt{5}%
})]=E_{bi}(\left| \Omega \right\rangle ).  \label{wav10}
\end{equation}
For any bipartition of $\left| \Omega \right\rangle $, the bipartition
relative entropy of entanglement $E_{bi}^R(\left| \Omega \right\rangle )$ is
just the entropy of the reduced density matrix, so $E_{bi}^R(\left| \Omega
\right\rangle )=E_{bi}(\left| \Omega \right\rangle ).$ However, the
tripartite relative entropy of entanglement $E_R(\left| \Omega \right\rangle
)$ should be no less than the bipartite one, as can be seen from the
definition of the relative entropy of entanglement. So we have
\begin{equation}
E_{bi}(\left| \Omega \right\rangle )=E_{MB}(\left| \Omega \right\rangle
)\geq E_R(\left| \Omega \right\rangle )\geq E_{bi}^R(\left| \Omega
\right\rangle ).
\end{equation}
So that all of them are equal for state $\left| \Omega \right\rangle .$ We
thus obtain the exact value of $E_{MB}(\left| \Omega \right\rangle )$ and
the tripartite relative entropy of entanglement $E_R(\left| \Omega
\right\rangle )$ for state $\left| \Omega \right\rangle .$

We may consider the minimal measurement entropy $E_{Hmin}$ defined in \cite
{Chernyavskiy} for $\left| \Omega \right\rangle $ state. The measurement
basis can be $\left| \phi _a\right\rangle \left| \phi _b\right\rangle
,\left| \phi _a\right\rangle $ $\left| \phi _b^{\perp }\right\rangle ,\left|
\phi _a^{\perp }\right\rangle \left| \phi _b\right\rangle ,\left| \phi
_a^{\perp }\right\rangle $ $\left| \phi _b^{\perp }\right\rangle .$ the
probabilities of the measurements are $p_{00}^{\prime }=\frac 15[1+xy],$ $%
p_{01}^{\prime }=\frac 15[1+x(1-y)],$ $p_{10}^{\prime }=\frac 15[1+(1-x)y],$
$p_{11}^{\prime }=\frac 15[1+(1-x)(1-y)].$ Where $x=(\left| a_0+a_1\right|
^2-\left| a_0\right| ^2)$ $\in [\frac 12(1-\sqrt{5}),$ $\frac 12(1+\sqrt{5}%
)],$ $y=(\left| b_0+b_1\right| ^2-\left| b_0\right| ^2)$ $\in [\frac 12(1-%
\sqrt{5}),$ $\frac 12(1+\sqrt{5})].$ Then the minimal entropy of the
measurement is
\begin{equation}
E_{Hmin}(\left| \Omega \right\rangle )=\min -\sum_{i,j=0}^1p_{ij}^{\prime
}\log _2p_{ij}^{\prime }=H_2[\frac 12(1+\frac 1{\sqrt{5}})].
\end{equation}
We can see that $E_{Hmin}(\left| \Omega \right\rangle )=E_{MB}(\left| \Omega
\right\rangle ).$

\subsection{Conditions for equal of EMB and minimal measurement entropy}

For a general tripartite state $\left| \psi \right\rangle $, we have $%
E_{Hmin}(\left| \psi \right\rangle )\geq E_{MB}(\left| \psi \right\rangle ),$
with the equality holds only when the basis $\left| \phi _b\right\rangle
,\left| \phi _b^{\perp }\right\rangle $ coincides with the basis $\left|
\phi _b^{\prime }\right\rangle ,\left| \phi _b^{\prime \perp }\right\rangle $%
. The basis $\left| \phi _b\right\rangle ,\left| \phi _b^{\perp
}\right\rangle $ are the eigenvectors of $B_0B_0^{\dagger }$, while the
basis $\left| \phi _b^{\prime }\right\rangle ,\left| \phi _b^{\prime \perp
}\right\rangle $ are the eigenvectors of $B_1B_1^{\dagger }.$ Hence only
when matrix $B_0B_0^{\dagger }$ commutes with $B_1B_1^{\dagger }$ can we
have $E_{Hmin}(\left| \psi \right\rangle )=E_{MB}(\left| \psi \right\rangle
).$

The $A_0$ and $A_1$ matrices for the standard form of tripartite state (\ref
{wav9}) are
\begin{equation}
A_0=\left[
\begin{array}{ll}
q_0 & 0 \\
0 & q_1
\end{array}
\right] ,\text{ }A_1=\left[
\begin{array}{ll}
0 & q_2 \\
q_3 & q_4e^{i\gamma }
\end{array}
\right] .
\end{equation}
Notice that $B_0B_0^{\dagger }+B_1B_1^{\dagger }=A_0^2+A_1A_1^{\dagger
}\equiv \mathcal{A},$ so the condition for the equality should be
\begin{equation}
\lbrack B_0B_0^{\dagger },\mathcal{A}]=0.  \label{wav11}
\end{equation}
If we require that $B_0B_0^{\dagger }$ commutes $B_1B_1^{\dagger }$ for all
measurements of the Alice's qubit, then condition (\ref{wav11}) reduces to $%
[A_0,A_1]=0,[A_0,A_1^{\dagger }]=0$ and $[A_1,A_1^{\dagger }]=0.$ These are
equivalent to $(q_0-q_1)q_2=0,(q_0-q_1)q_3=0,$ $q_2=q_3,q_2e^{i\gamma
}=q_3e^{-i\gamma },q_2e^{i\gamma }=q_3e^{-i\gamma }.$ The solutions should
be either
\begin{equation}
q_0=q_1,q_2=q_3,\gamma =0,
\end{equation}
or
\begin{equation}
q_2=q_3=0.
\end{equation}
The corresponding states are
\begin{eqnarray*}
\left| \Omega _1\right\rangle &=&q_0(\left| 000\right\rangle +\left|
011\right\rangle )+q_2(\left| 101\right\rangle +\left| 110\right\rangle
)+q_4\left| 111\right\rangle , \\
\left| \Omega _2\right\rangle &=&q_0\left| 000\right\rangle +q_1\left|
011\right\rangle +q_4e^{i\gamma }\left| 111\right\rangle .
\end{eqnarray*}

For $\left| \Omega _1\right\rangle $ state, we choose $B_0=\cos \theta
A_0+\sin \theta A_1.$ Let $\det (B_0)=0$ to determine $\theta ,$ then the
eigenvalues of $B_0B_0^{\dagger }$ are $0$ and $(TrB_0)^2.$ We have
\begin{equation}
E_{MB}(\left| \Omega _1\right\rangle )=E_{bi}(\left| \Omega _1\right\rangle
)=H_2[\frac 12(1+\sqrt{1-C^2})],
\end{equation}
with $C^2=4q_0^2[2(1-2q_0^2)-q_4^2].$ The tripartite relative entropy of
entanglement $E_R(\left| \Omega _1\right\rangle )$ is obtained to be equal
to $E_{bi}(\left| \Omega _1\right\rangle )$ since it is in between $%
E_{MB}(\left| \Omega _1\right\rangle )$ and $E_{bi}(\left| \Omega
_1\right\rangle ).$ Similar results can be obtained for states $\left|
\Omega _1^{\prime }\right\rangle =q_0(\left| 000\right\rangle +\left|
101\right\rangle )+q_3(\left| 011\right\rangle +\left| 110\right\rangle
)+q_4\left| 111\right\rangle $ and $\left| \Omega _1^{\prime \prime
}\right\rangle =q_0(\left| 000\right\rangle +\left| 110\right\rangle
)+q_1(\left| 101\right\rangle +\left| 011\right\rangle )+q_4\left|
111\right\rangle .$

For $\left| \Omega _2\right\rangle $ state, the equality of EMB and the
bipartite entanglement do not hold in general, however, when $q_1=0,$ we do
have the equality. But the situation seems rather trivial.

\subsection{LOCC monotone for completely measurement of a pure tripartite
state}

A fundamental property of an entanglement measure is that it should not
increase under LOCC. Local measurement will not increase the entanglement of
a state on average. To illustrate the detail meanings of EMB under LOCC,
let's consider the tripartite qubit state first. Given a pure tripartite
qubit state (\ref{wav6}) with coefficients $A_{ijk},$ we can calculate the
bound with formula (\ref{wav7}) where the default first step measurement is
on Alice's qubit. We may first measure Bob's qubit or Charlie's qubit. The
results may differ. The bound should be the minimum of the three by
definition. We denote it as
\[
E_{MB}(\left| \psi \right\rangle )=\min \{E_{MB}^A(\left| \psi \right\rangle
),E_{MB}^B(\left| \psi \right\rangle ),E_{MB}^C(\left| \psi \right\rangle
)\},
\]
where $E_{MB}^i(\left| \psi \right\rangle )$ is calculated with formula (\ref
{wav7}) when $ith$ partite is measured first. One the other hand, after a
measurement on Alice's partite, the state left should be (\ref{wav7a}) with
probability (\ref{wav7c}) or (\ref{wav7b}) with probability (\ref{wav7d}).
The maximal average entanglement after local measurement on Alice's partite
can be denoted as
\begin{equation}
E_{LOCC}^A(\left| \psi \right\rangle
)=\max_{a_0,a_1}[-\sum_{i=0}^1p_iE(\left| \psi _i\right\rangle )].
\end{equation}
We may measure Bob's or Charlie's qubit first, the maximal average
entanglement after a local measurement then is
\begin{eqnarray}
E_{LOCC}(\left| \psi \right\rangle ) &=&\max \{E_{LOCC}^A(\left| \psi
\right\rangle ),  \nonumber \\
&&E_{LOCC}^B(\left| \psi \right\rangle ),E_{LOCC}^C(\left| \psi
\right\rangle )\}.
\end{eqnarray}
If we have
\begin{equation}
E_{MB}(\left| \psi \right\rangle )\geq E_{LOCC}(\left| \psi \right\rangle ),
\label{wav12}
\end{equation}
then the EMB is an LOCC monotone, we may call it measurement entanglement
and denoted as $E_M(\left| \psi \right\rangle ).$ In the following we will
prove that (\ref{wav12}) is true for a pure tripartite state in the sense of
completely measurement of the first partite.

\begin{theorem}
Entanglement measurement bound for a pure tripartite qubit state is an LOCC
monotone.
\end{theorem}

Proof: Suppose that EMB of a tripartite pure state $\left| \psi
\right\rangle $ is achieved by measuring Alice's partite first, then we have
\begin{equation}
E_{MB}(\left| \psi \right\rangle )=E_{MB}^{(A)}(\left| \psi \right\rangle
)\geq E^{(AB,C)}(\left| \psi \right\rangle )
\end{equation}
by (\ref{wav5}), where $E^{(AB,C)}(\left| \psi \right\rangle )$ is the
bipartite entanglement. When we measure on Alice or Bob of $AB$ part, the
average entanglement of the remained part will not exceed $E^{(AB,C)}(\left|
\psi \right\rangle )$ according to the monotonicity of bipartite
entanglement \cite{Bennett}, namely,
\begin{eqnarray*}
E^{(AB,C)}(\left| \psi \right\rangle ) &\geq &E_{LOCC}^A(\left| \psi
\right\rangle ), \\
E^{(AB,C)}(\left| \psi \right\rangle ) &\geq &E_{LOCC}^B(\left| \psi
\right\rangle ).
\end{eqnarray*}
Similarly, we also have $E_{MB}^{(A)}(\left| \psi \right\rangle )\geq
E^{(B,AC)}(\left| \psi \right\rangle )$ and the monotonicity of bipartite
entanglement shows that $E^{(B,AC)}(\left| \psi \right\rangle )\geq
E_{LOCC}^C(\left| \psi \right\rangle ).$ Thus (\ref{wav12}) is proved, and
the theorem follows.

For a $d_1\times d_2\times d_3$ tripartite state with completely measurement
of the each partite, we have
\[
E_M(\left| \psi \right\rangle )=E_{MB}(\left| \psi \right\rangle )\geq
E_{LOCC}(\left| \psi \right\rangle ).
\]
The completely measurement means that the state of $N$ parties is projected
to $N-1$ parties after the measurement.

\section{Conclusion}

The entanglement bound based on local measurements is introduced for
multipartite pure states. The measurement sequence is a dependent one, for
each step of measurement, the basis rely on the former measurement results.
The entanglement measurement bound defined in this paper is a lower bound of
a multipartite entanglement measure called \textit{minimal measurement
entropy }which is based on independent measurements of the parties. The
entanglement measurement bound is also the upper bound of \textit{geometric
measure }and \textit{the relative entropy of entanglement. }The property of
coarser grain for the bound is derived. Based on the coarser grain of the
bound and the fact that in bipartite case the bound is equal to the relative
entropy of entanglement, we obtain the lower and upper bounds for the
relative entropy of entanglement of a tripartite state. For a tripartite
qubit state we derive the condition when the lower and upper bound coincide.
The exact relative entropy of entanglement follows for a class of tripartite
qubit states in the form of $\left| \Omega _1\right\rangle =q_0(\left|
000\right\rangle +\left| 011\right\rangle )+q_2(\left| 101\right\rangle
+\left| 110\right\rangle )+q_4\left| 111\right\rangle $ or their qubit
permutation states. It is an interesting phenomenon that the tripartite
relative entropy of entanglement is equal to the bipartite relative entropy
of entanglement while the tangle is nonzero for these states. For tripartite
qubit states, the bound itself is an entanglement monotone. Further works
can be done on whether the bound is an LOCC monotone or not in general.

.

\section*{Acknowledgment}

Funding by the National Natural Science Foundation of China (Grant No.
60972071), Natural Science Foundation of Zhejiang Province (Grant No.
Y6100421), Zhejiang Province Science and Technology Project (Grant No.
2009C31060), Zhejiang Province Higher Education Bureau Program (Grant No.
Y200906669) are gratefully acknowledged.

\end{document}